# A Compact Verilog-A ReRAM Switching Model

Ioannis Messaris *Student Member, IEEE*, Alexander Serb *Member, IEEE*, Ali Khiat, Spyridon Nikolaidis *Senior Member, IEEE,* Themistoklis Prodromakis *Senior Member IEEE*

*Abstract—* The translation of emerging application concepts that exploit Resistive Random Access Memory (ReRAM) into large-scale practical systems requires realistic, yet computationally efficient, empirical models that can capture all observed physical devices. Here, we present a Verilog-A ReRAM model built upon experimental routines performed on *TiOₓ*-based prototypes. This model was based on custom biasing protocols, specifically designed to reveal device switching rate dependencies on a) bias voltage and b) initial resistive state. Our model is based on the assumption that a stationary switching rate surface $m(R,v)$ exists for sufficiently low voltage stimulation. The proposed model comes in compact form as it is expressed by a simple voltage dependent exponential function multiplied with a voltage and initial resistive state dependent second order polynomial expression, which makes it suitable for fast and/or large-scale simulations.

## I. Introduction

SINCE 2008 when the basic resistive switching property of a double-layer nano-scale film based on Titanium dioxide was studied [1] and linked to Chua's theory of the 'memristor' [2], understanding of practical memristor realisations has moved far beyond the simple 'moving barrier' model. Solid-state memristor devices stem from different technological roots (phase-change memory, spin-torque transfer, metal-oxide etc. [3], [4], [5] ) and employ a variety of electrode/active layer materials and geometries. Such devices are becoming more and more accessible to researches, and it is now clear that each implementation features properties that render them suitable for different applications. There are memristors that have been reported to switch quickly and in a probabilistic fashion [6], while others can have their resistive state (RS) shifted in small steps (analogue mode) [7] which is ideal for synaptic learning [8].

Resistive Random Access Memory (ReRAM) devices have perhaps received the largest attention as they support multi-state programming, can be programmed swiftly and with

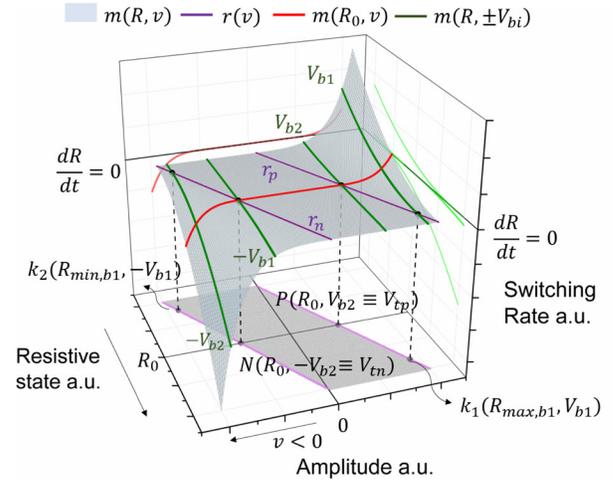

Fig. 1. (a) Example switching rate surface ( $m(R,v)$ - gray surface) characteristics as reproduced by the proposed model (8). Green lines correspond to $dR/dt$ vs $R$ plots for constant bias stimulation ( $V_b$ ) ($m(R,\pm V_{bi})$) while red line captures $dR/dt$ dependence on bias voltage ($v$) when switching from the same initial RS ($R_0$) ($m(R_0,v)$). Purple lines delineate the 'absolute voltage threshold' function and are projected on the $R-v$ plane to mark the interface between finite non-zero (white) and zero (gray) switching regions. Points $N$ and $P$ pinpoint the absolute voltage threshold points exhibited when the device is at $R = R_0$ for both voltage polarities. Conversely, the purple lines can also be interpreted as the 'max/min resistive state for which a fixed bias voltage is effective. For example, points $k_1$ and $k_2$ mark the max. RS for which stimulation at $V = V_{b1}$ is effective, similar for mix. RS and $V = -V_{b2}$.

low energy and can be compatible with post-CMOS processing. As ReRAM technology matures, a need develops for: a) automated testing routines intended to accelerate the process development cycle [7], [9] and b) robust computer models enabling informed memristor-based system design before committing to silicon. To date, several empirical and semi-empirical models have been published accounting for either only non-volatile [10], [11] or both volatile and non-volatile behaviours [12], [13]. The majority of the published compact memristor models are SPICE models and although helpful guidelines for accurate and reliable modelling of memristors in the SPICE environment have been published [14], the fact remains that such models lack industrial multi-simulator compatibility. On the other hand, Verilog-A is a behavioural language promoted by the Compact Model Council and thus has emerged as the de facto standard language for defining and distributing compact models for both academic and industrial research groups, mainly due to its flexibility to run in numerous industrial electrical simulators (Spectre,

This work was supported in part by the EU COST Action IC1401 "MEMOCIS" and the Engineering and Physical Sciences Research Council under Grant EP/K017829/1.

The data from this paper can be obtained from University of Southampton e-Prints Repository DOI:10.5258/SOTON/403321.

Ioannis Messaris and Spyridon Nikolaidis are with the Department of Physics, Aristotle University of Thessaloniki, 54124 Thessaloniki Greece (email: imessa@physics.auth.gr).

Alexander Serb, Ali Khiat and Themis Prodromakis are with the Nano Group, ECS, University of Southampton, Highfield, Southampton SO17 1BJ. (email: A.Serb@soton.ac.uk).

HSPICE, ADS, Eldo etc.).

In this work, we employ two testing biasing routines specifically designed for characterising $TiO_2$-based ReRAM prototypes in an operationally relevant environment and use the resulting measured data sets in order to model the technology. Data from these routines is aggregated to create a data-driven ReRAM model that captures the switching rate ($dR/dt$) dependency on both RS and bias conditions ($R,v$), i.e. reconstructs the $\Delta R(R,v)$ surface (assumed to be stationary for sufficiently low voltage biasing) (Fig. 1). The resulting model combines a simple voltage controlled exponential expression $s(v)$ with a second order voltage and RS dependent polynomial expression $f(R,v)$ thus is suitable for integration in circuit simulators.

Section II describes model functionality, operating principles and explains the experimental algorithms performed on the Device Under Test (DUT). Section III specifies the processing of the extracted measurements to the end of producing meaningful data. This data is then exploited for the extraction of the proper parameter values that fit the proposed model to the DUT. Section IV: a) approximates the extracted model with continuous and differential expressions and b) reveals crucial coding details adopted in the proposed Verilog-A implementation. Finally, section VI concludes the paper.

## II. MODEL AND METHODS

This section focuses on revealing the operating principles of the proposed model. These result from targeted experimental testing which is also described. We demonstrate the modeled RS switching rate expression which is based on the following concepts and assumptions: a) switching sensitivity is a stationary function of initial RS and bias voltage $\Delta R(R,v)$, b) the switching rate surface ($dR/dt = m(R,v)$) can be approximated by a product of the form: $s(v) \times f(R,v)$, where the terms correspond to the functional forms of the 'switching sensitivity' function $s(v)$ and the 'window function' $f(R,v)$ and c) finite and zero switching resistive regions are separated by the presence of an absolute threshold curve that traverses the $R-v$ plane. This is defined by the 'absolute voltage threshold' function $r(v)$ that is nested in $f$ and essentially regulates its windowing behavior.

The stated assumptions are validated in section III where functions $s$, $f$ and $r$ are derived from measurements of switching ($\Delta R$) recorded using the described biasing schemes on an in-house $TiO_x$-based sample.

Importantly, throughout this work, device RS is formally defined as static resistance at a standardized read-out voltage. Regardless of the voltage used to bias the device for the purposes of switching, all assessments of RS are carried out at the standard voltage ($0.2V$). Thus the model is designed to describe changes in RS as measured by a standardized way in response to input stimulation.

### A. The switching rate model expressions

Experimentally, device RS is defined as:

$$R = V/I \text{ at } V = V_{read}, \tag{1}$$

where $V = V_{read}$ is the read voltage applied and $I$ is the current flowing through device terminals during read-out.

We present an empirical, RS and voltage controlled switching rate memristor model ($dR/dt$) that describes changes in RS as a result of stimulation given at a well-defined RS $R_0$ and with a fixed voltage $v$:

$$\frac{dR}{dt} = m(R,v) = s(v)f(R,r(v)), \tag{2}$$

where, $m(R,v)$ manifests the three dimensional surface that tracks device switching rate as a function of initial RS ($R_0$) and bias voltage application ($v$) (Fig. 1). Function $s(v)$ corresponds to the 'switching sensitivity' and is solely voltage controlled expressed in units $s^{-1}\Omega^{-1}$. Function $f(R,v)$, the device Window Function (WF), is both RS and voltage dependent measured in units $\Omega^2$ with its specific voltage dependency regulated by the nested 'absolute voltage threshold' function $r(v)$. What is interesting with $r$ is that it is measured in $\Omega$ and thus is found on the $R-v$ plane. Its physical interpretation is of particular importance as it models the device absolute threshold curve (plotted on the $R-v$ plane) that separates finite from zero switching rate regions (Fig. 1). Importantly, all functions in (2) have a piecewise nature where for each one, positive and negative stimulation branches have the same functional forms but with different parameter values.

### B. Experimental testing

Fig. 1 displays a sample switching rate surface that resembles the form of the experimentally measured data according to which $m(R,v)$ is modeled in Section III (Fig. 7). In order to reconstruct $m$ from measured results the DUT is subjected to two distinct experimental protocols:

1) *The 'biasing optimizer'*

   Pulsed voltage ramps of alternating polarities are employed and the DUT reacts by oscillating its RS around an initial value $R_0$ revealing the relation between $\Delta R$ and bias voltage around $R_0$ $s(v)|_{R=R_0} \propto \Delta R(R_0,v)$ (the 'switching sensitivity' function). Each ramp level is a pulse train that consists of $N\ pulses/train$ with fixed duration $t_w$. The observed, bipolar DUT behaviour exhibiting SET transitions under positive voltage bias and RESET under negative is exploited by the algorithm in order to restrict DUT RS within a narrow resistive range. Pulse train polarity is changed each time DUT RS exits a user defined tolerance band $\varepsilon_{opt}$ (in % of $R_0$) around $R_0$. The result is an estimate of the switching rate function around $R_0$ ($m(R_0,v)$). For both routines, $\Delta R$ measurements are converted into a switching rate ($dR/dt$) by dividing $\Delta R$ with the duration of the stimulus [7].

2) *The 'operating RS sweeper'*

   This algorithm applies a train consisting of $S$ identical pulses of fixed pulse duration $t_w$. The DUT reacts by changing its RS and the relation between switching $\Delta R$ and running $R$ is revealed at the chosen bias voltage $f(R)|_{v=V_b} \propto \Delta R(R, V_b)$ (device WF). This is performed for multiple voltage levels $V_b$. In its flow, the algorithm is carried out by changing polarity with each applied train and increasing the amplitude every two trains by a defined step voltage $V_{step}$. The absolute amplitude of the pulses applied will scale according to $V_{step}$ from an initial voltage $V_{start}$, up to a user defined value $V_{stop}$ and the algorithm will terminate after pulse trains featuring this value in both polarities are applied.

These experimental routines are designed to isolate the $R$ and $v$ dependencies by slicing the $m(R, v)$ surface parallel to the RS ($R$) (sweeper) and voltage ($v$) (optimizer) axes respectively. In Fig. 1, red line corresponds to the $dR/dt - v$ plot for fixed initial $R_0$ ($m(R_0, v)$) while green lines capture $dR/dt - R$ dependency for constant bias voltage $\pm V_{b1,2}$ ($m(R, \pm V_{b1,2})$). Both testing routines where implemented on a general purpose ReRAM characterisation instrument previously described in [15], [16] in order to characterise a $TiO_x$ sample, cell of a stand-alone ReRAM device array, described in [5]. The specific experimental routines employed on the modeled DUT are shown in Fig. 5: (a) corresponds to the 'optimizer' routine while (b) to the 'sweeper'. The parameter values of the testing schemes designed to characterize the DUT along with the data processing of the exported data towards gaining meaningful resistive switching information are described in section III.

### C. Switching rate surface properties

Fig. 2(a) plots a family of sample switching rate functions $m(R, V_b)$ for various bias voltages $V_b$ of both polarities, where device RS is swept in its resistive region of operation. The forms of the illustrated curves follow the forms of the plots fitted to the corresponding experimental data (Fig. 3 (a-c)) measured from the DUT. We notice bipolar switching, where higher absolute voltage amplitudes provoke higher switching rates and switching intensifies as we move away from the resistive range boundaries $R_{min}$ and $R_{max}$ for constant voltage stimulation. We also see that switching beyond these boundaries is set to zero (for the corresponding bias voltage amplitude) i.e. the plots in Fig. 2(a) embody device WF characteristics. Furthermore, the positions of these resistive limits are monotonically dependent on voltage. Higher negative voltages push $R_{min}$ to lower resistive values while more invasive positive biases push $R_{max}$ to higher RSs. The physical interpretation of these boundaries is that for any RS below $R_{max}$ (active region) at a given positive voltage, applying this voltage can push the device to $R_{max}$, but no further (saturation). If the device is already above $R_{max}$, no switching occurs (for positive stimulation). The same applies for negative biasing and the $R_{min}$ limit.

Fig. 2(b) illustrates the corresponding to Fig. 2(a), family of switching rate plots $m(R_0, v)$ starting from different initial RSs ($R_0$) and sweeping the applied voltage ($v$) for each case. Again, the curve forms shown in this figure resemble the corresponding experimental data demonstrated in Fig. 4(d).

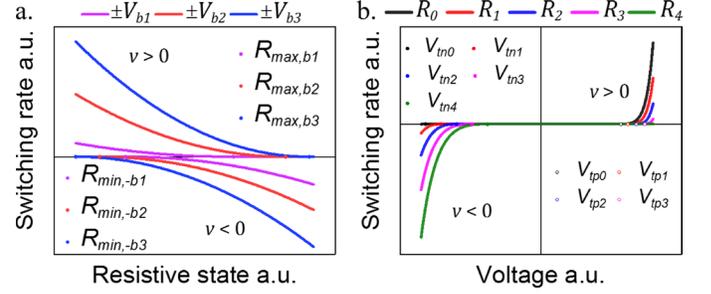

Fig. 2. Example switching rate plots for (a) $dR/dt$ vs $R$ for three voltages where $|V_{b1}| < |V_{b2}| < |V_{b3}|$ and (b) $dR/dt$ vs $v$ for various initial RSs for which $R_0 < R_1 < R_2 < R_3 < R_4$ Round symbols on the $x$-axis correspond to the absolute threshold values which are resistive for (a) and voltages for (b).

Stronger voltage biases cause higher switching rates, as does moving away from the resistive range boundaries of operation, i.e. towards lower RS values for $v > 0$ and higher RS values for $v < 0$. This is consistent with the $m(R, V_b)$ plots of Fig. 2(a).

What is particularly notable in the presented characteristic plots is that Fig. 2(a) exhibits resistive thresholds while Fig. 2(b) shows voltage thresholds. In Fig. 1 the two purple lines (for the positive and negative stimulation cases) delineate the zero switching surface $dR/dt = z(R, v) = 0$ (projected on the base of Fig. 1 as the grey area), where $z$ defines a bijective function between $R$ and $v$ (which as mentioned are monotonically linked) for which voltage and resistive thresholds are given by $V_{tpn} = r^{-1}(R)$ and $R_{min/max} = V^{-1}(v)$ respectively. These boundaries are modeled in (2) by the 'absolute voltage threshold' function $r(v)$, where $r_p(v)$ and $r_n(v)$ correspond to the positive and negative branches respectively. The switching rate plots $m(R, V_{b1})$ and $m(R, -V_{b1})$ (green lines) represent the plots of Fig. 2(a) and cross $r$ at points $k_1(R_{max,b}, V_{b1})$ and $k_2(R_{min,b}, -V_{b1})$, thus defining the values of the resistive thresholds ($R_{max,b}, R_{min,b}$). Similarly, the red line ($m(R_0, v)$), which exemplifies the switching rate plot family of Fig. 2(b), crosses $r_p$ and $r_n$ at points $P(R_0, V_{tp})$ and $N(R_0, V_{tn})$, where $V_{tp}$ and $V_{tn}$ are the positive and negative voltage thresholds that correspond to device RS equal to $R_0$. According to the bijective property of $z$, points $P$ and $N$ serve also as resistive threshold points for the RS dependent switching rate plots $m(R, V_{b2} \equiv V_{tp})$ and $m(R, -V_{b2} \equiv V_{tn})$ (the green curves in Fig. 1).

### D. Functional forms

Next, we define the mathematical forms of the functions comprising (2). In order to fit the experimental switching rate plots of Fig. 3(a-c) for constant voltage bias $V_b$, we seek for an RS dependent expression that matches the demonstrated data and is formulated in a way so as to model zero switching at a defined resistive threshold value. Accordingly, they can be described with an RS dependent second order law of the form,

$$\frac{dR}{dt}\bigg|_{V_b} = \begin{cases} s_p(V_b)\big(r_p(V_b) - R\big)^2 stp(r_p(v) - R), V_b > 0, R < r_p(v) \\ s_n(V_b)\big(R - r_n(V_b)\big)^2 stp(R - r_n(V_b)), V_b \leq 0, R \geq r_n(v) \end{cases},$$
(4)

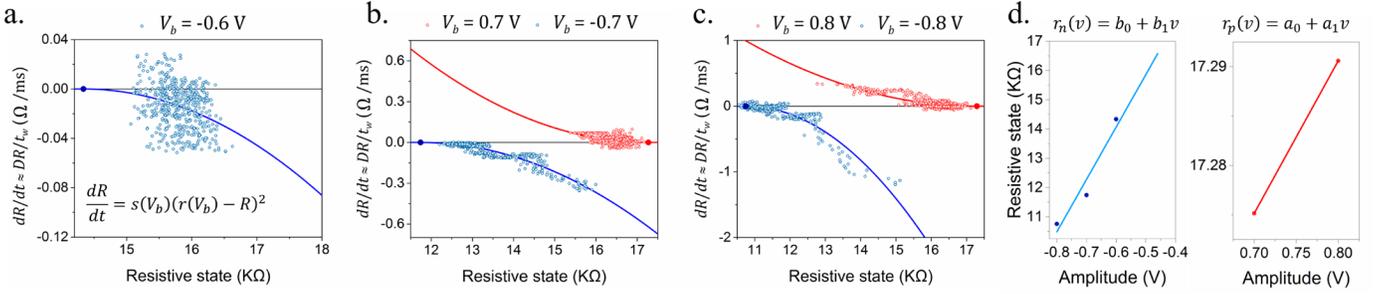

Fig. 3. (a-c) Red and blue symbols correspond to experimental $dR/dt$ vs $R$ points after suitable processing of the data exported by the sweeper testing routine. Red and blue lines are reproduced by the $2^{nd}$ order expression (4). Filled dots indicate the resistive boundaries of operation beyond which switching is nullified ($R_{min/max}$). Three absolute voltage levels are investigated, 0.6V, 0.7V and 0.8V. The positive stimulation case for 0.6V is omitted as the obtained data for this voltage level was too noisy to provide meaningful information. (d) Linear fitting of the resistive boundary values revealed after fitting (4) to the switching rate data shown in (a-c). Left panel: Red symbols are the calculated $R_{min}$ points while red line is reproduced by $r_n(v)$ (5). Right panel: Blue symbols correspond to the $R_{max}$ points while blue line is exhibited by $r_p(v)$ (5).

where $R$ is device RS, $f(R,v)$ is the device WF and $s_{p/n}(V_b)$, $r_{p/n}(V_b) = R_{max/min V_b}$, are the switching sensitivity and resistive boundary values. Subscripts $p$ and $n$ denote the positive and negative stimulation cases respectively. The latter ($r_{p/n}$) determine the device resistive range of operation for bias voltage $V_b$. Practically the squared parenthetic term ensures zero switching at the resistive threshold point determined by $r(v)$, while the step function $stp(\cdot)$ imposes zero switching beyond this point. Fig. 3(a-c) also demonstrates the voltage dependent nature of the $R_{min/max}$ values modeled by the 'absolute threshold' function $r(v)$ and switching activity that coincides with the conditionals in (4). The behavior of the switching sensitivity function $s_{p/n}$ will be examined towards the end of this section.

The expression that fits the zero switching boundary values $R_{min/max}$ is a first order expression (Fig. 3(d)), thus defining the form of the $r(v)$ function in (2) as:

$$r(v) = \begin{cases} r_p(v) = a_0 + a_1 v, & v > 0 \\ r_n(v) = b_0 + b_1 v, & v \leq 0 \end{cases}, \quad (5)$$

where $a_0, a_1, b_0, b_1$ are fitting parameters. Higher positive biases ($v > 0$) push $R_{max}$ from its minimum value $a_0$ (for $v = 0$) to higher resistive values, while more invasive negative biases ($v < 0$) push the highest exhibited $R_{min}$ value $b_0$ (for $v = 0$) to lower resistive values. Parameters $a_1$ and $b_1$ define the rate at which the boundary resistive values change proportionally to the applied voltage.

Solving the equations in (5), with respect to voltage gives us the switching threshold voltages $V_{tpn}$ for which an RS on the $r(v)$ curve becomes a resistive range boundary:

$$V_t(R) = \begin{cases} V_{tn}(R) = (R - a_0)/a_1, & v > 0 \\ V_{tp}(R) = (R - b_0)/b_1, & v < 0 \end{cases}, \quad (6)$$

We proceed on defining the form of the 'switching sensitivity' function $s(v)$ by investigating DUT switching dependency on biasing conditions when stimulated from the same initial RS $R_0$. As $f$ and $r$ are given in (4), (5), only $s$ is left to be determined for the complete functional form of the switching rate function $m(R,v)$ to be revealed. Expression (2) combined with (4), (5) fits the corresponding experimental results shown in Fig. 4(d) when $s$ takes the form of a simple exponential function defined as:

$$s(v) = \begin{cases} A_p\left(-1 + e^{\frac{|v|}{t_p}}\right), & v > 0 \\ A_n\left(-1 + e^{\frac{|v|}{t_n}}\right), & v \leq 0 \end{cases}, \quad (7)$$

where $A_p, A_n, t_p, t_n$ are fitting parameters. The simple form of (7) renders the device RS to be idle in the absence of voltage stimulation ($s(0) = 0 \Rightarrow m(R,v) = 0$) and therefore ensures non-volatile behavior for the model.

The final model expression results from combining (2), (4), (5) and (7). The proposed RS and voltage dependent switching rate function $m(R,v)$ governing the device is expressed as:

$$\frac{dR}{dt} = m(R,v) =$$
$$A_p\left(-1 + e^{\frac{|v|}{t_p}}\right)(r_p(v) - R)^2 stp(r_p(v) - R) stp(v) +$$
$$A_n\left(-1 + e^{\frac{|v|}{t_n}}\right)(R - r_n(v))^2 stp(R - r_n(v)) stp(-v), \quad (8)$$

where $r_p, r_n$ are those defined in (5), and $A_p, A_n, t_p, t_n, a_0, a_1, b_0, b_1$ are fitting parameters.

III. RESULTS

We proceed by specifying the parameters for the experimental testing and using the resulting switching data ($\Delta R$) to extract the model parameter values that fit (8) to the DUT, i.e. our modelling target.

A. 'Optimizer' parameters and data refinement

The optimizer routine run employed on the DUT is shown in Fig. 5(a). The routine parameter values are, $N = 10 \, pulses/train$, $t_w = 100 \mu s/pulse$ and $\varepsilon_{opt} = 10\%$. The result is an estimate of the switching rate function around $R_0 = 13.65 K\Omega$ ($m(R_0, v)$) (Fig. 5(a)).

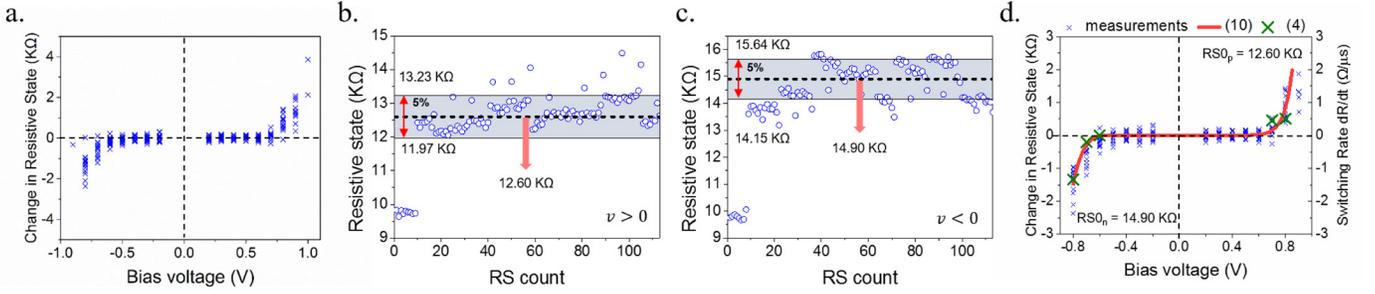

Fig. 4. (a) Change in RS recorded after application of each train of programming pulses as a function of pulse voltage ($\Delta R(v)$). (b) and (c): Application of the refinement algorithm on 'optimizer' measurements that correspond to positive (b) and negative (c) pulse stimulation. Grey shadings highlight the most data-populated resistive bands along with the band's resistive limits and the corresponding tolerance $\varepsilon_{ref}$ according to which they are defined. (d) Left y-axis: Symbols correspond to change in RS ($\Delta R$) recorded after refining measurements, vs bias voltage $V_b$. Red line is reproduced by the proposed model (8). The reference RS for positive ($RS0_p = 12.60K\Omega$) and negative ($RS0_n = 14.90K\Omega$) stimulation is also shown. Right y-axis: RS switching rate versus bias voltage $V_b$. Green symbols are switching rate points calculated by (4), for the fitted $k_n(V_b), k_p(V_b), r_n(V_b), r_p(V_b)$ for the sweeper voltages $V_b = \{-0.6V, \pm 0.7V, \pm 0.8V\}$ (0.6V case is neglected). As (4) results solely from sweeper data, the allocation of these points corroborate device stability hypothesis.

Naturally, the optimiser process causes excursions of DUT RS away from $R_0$, so the estimate of $m(R_0, v)$ is imperfect. To mitigate this effect, data extracted via the 'optimizer' could be filtered so that $m(R_0, v)$ is estimated using only data for $R \in [R_0 - \varepsilon_{ref}, R_0 + \varepsilon_{ref}]$ for some chosen $\varepsilon_{ref}$ (expressed as % of $R_0$). This data refinement process helps define an RR range that is both narrow enough to be 'sufficiently close' to its centre RS and yet at the same time adequately populated with measurements to offer useful information.

However, in practice most of the data $\Delta R(R, v)$ is not always clustered around the optimiser's operating $R_0$. Therefore, it is more useful to determine the value of $R_0$ a posteriori, i.e. after the completion of the optimiser run. The $R_0$ around which most gathered data is clustered can be defined as the central value of $R_x$ for which the interval $[R_x - \varepsilon_{ref}, R_x + \varepsilon_{ref}]$ contains the maximum number of data points. We then consider that the results of the optimiser trace a line in the $R - V$ plane at $\Delta R(R_x, v)$ instead of $\Delta R(R_0, v)$ as a fairer approximation. Fig. 4(b, c) illustrates the refinement process as applied on positive and negative stimulation data for $\varepsilon_{ref} = 5\%$. For (b), the most populated interval has its centre at $12.60K\Omega$ while for (c), the most populated interval has its centre at $14.90K\Omega$, even though according to the optimiser the central value of $R_0$ should be $13.65K\Omega$ (Fig. 5(a)). The resulting refined 'switching' plot is shown in Fig. 4(d). Finally, amount of switching data ($\Delta R$) is converted into switching rate data ($dR/dt$): $\Delta R/(N \times t_w) = \Delta R/1ms$.

B. *'Sweeper' parameters and data processing*

Fig. 5(b) shows the sweeper test run results on the same DUT. Parameter values are $S = 500 \; pulses/train$, $t_w = 100\mu s/pulse$, $V_{step} = 0.1V$, $V_{start} = 0.6V$ and $V_{stop} = 0.8V$.

The resulting information linking RS switching to stimulation voltage seen in Fig. 5(b) is processed by converting $RS(pulse)$ into $RS(t)$ by taking into account the amplitude ($V_b$) and duration ($t_w$) of the voltage pulses. Subsequently, a smoothing time derivative of the RS measurements is employed and the switching rate, $dR/dt$ versus RS, $R$ for constant voltage application, $V_b$ is captured in Fig. 3(a-c).

Besides DUT RS sensitivity, Fig. 3(a-c) also shows the dependency of $R_{min/max}$ on voltage stimulation.

It is important to note that timing between successive reads and successive writes is not under user control in the current 'optimizer' and 'sweeper' implementations. Instead, pulses are sent as soon as the system is ready to source them (system automatically determines how long to measure in order to obtain a reliable measurement). Timing choices for these actions may affect DUT behaviour and results (e.g. timing between writes may affect DUT behaviour through possible thermal effects [13] and timing between reads will affect algorithm results in samples exhibiting noticeable RS volatility [17]). These effects are subjects of further, ongoing work.

C. *Model parameter extraction for the DUT*

In this sub-section we fit the proposed model expression (8) to the corresponding data conceived by the presented device testing schemes.

*1) Window function f*

The switching rate plots of Fig. 3(a-c) are fitted quite well with the suitable second order expression (4) where $s_{n/p}(V_b)$ and $r_{p/n}(V_b) = R_{max/min}$ reduce to constant values when $V_b$ is fixed.

Fig. 3(d) plots the $R_{min/max}$ points that result after fitting the switching rate data of Fig. 3(a-c). The linear expressions in (5) are used to express the resistive boundary point trends thus defining the relevant parameter values for the specific DUT as:

$$a_0 = 17.16K\Omega \;, \; a_1 = 0.15K\Omega/v \;, \; b_0 = 24.81K\Omega \;, \; b_1 = 17.91K\Omega/v \quad (9)$$

We notice that the DUT exhibits rather stable $R_{max}$ points in contrast to $R_{min}$ which, depending on the voltage applied, span to the entire DUT range of operation ($10K\Omega - 17K\Omega$, Fig. 3(d)).

*2) Switching sensitivity function s*

We use (8) for $v > 0$ to fit the positive branch of the refined switching rate data (Fig. 4(d)) for $R = RS0_p = 12.60K\Omega$ and the corresponding parameter values in (9) ($a_0$, $a_1$).

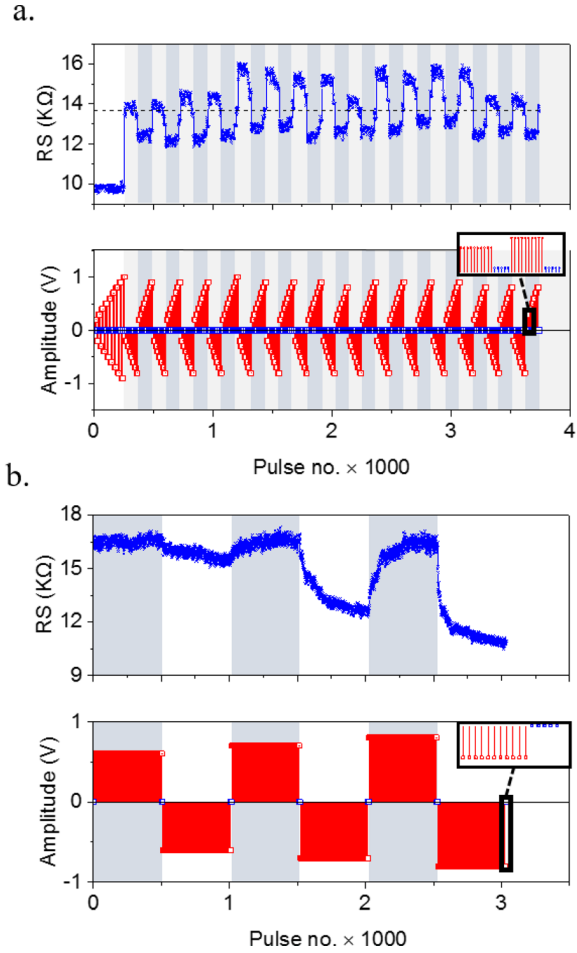

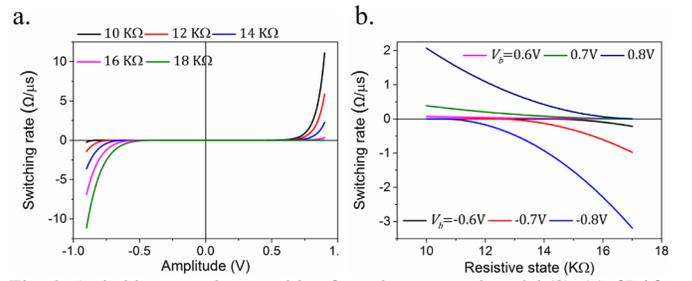

Fig. 6. Switching rate plots resulting from the proposed model (8). (a) $dR/dt$ vs $v$ for initial RSs that belong to the RR the DUT visits during our experiments ($10K\Omega - 18K\Omega$). (b) $dR/dt$ vs $R$ for the three absolute voltage levels employed by the sweeper testing routine ($V_b = \pm(0.6V - 0.8V)$).

Fig. 5. (a) Optimizer test results and (b) Sweeper test results. Both display evolution of measured DUT RS with input programming pulse sequence. Top traces for each: RS measurements. Bottom traces: Programming pulse sequences. Gray striations in (a) and (b) separate SET transitions under positive voltage bias and RESET under negative. The $R_0 = 13.65K\Omega$ value in (a) denotes the RS around witch the device cycles under the optimizer routine.

Analogously, we fit the negative branch for $R = RS0_n = 14.90K\Omega$ and the $b_0$, $b_1$ parameter values in (9). The two discreet fittings reveal the values of the remaining parameters $A_p$, $A_n$, $t_p$ and $t_n$:

$A_p = -4.86 \times 10^{-5} \Omega^{-1} s^{-1}$ , $t_p = 0.12V$ , $A_n = 1.09 \times 10^{-3} \Omega^{-1} s^{-1}$, $t_n = 0.18V$ (10)

*3) Combining f and s.*

Experimentally we find that $f$ and $s$ are mutually consistent: Fig. 4(d), plots experimental $dR/dt$ points versus $v$ ($f$) for the same initial, $R_0$. Green $x$-points show corresponding $dR/dt$ values resulting from (4) (Fig. 3(a-c)) along with the parameter values (9) for $R = RS0_p = 12.60K\Omega$ and $R = RS0_n = 14.90K\Omega$. We notice that the resulting points are consistent with the refined switching rate measurements therefore supporting the device response stability hypothesis. Fig. 6 displays simulated switching rate plot families that correlate to the modeled DUT for a) constant biasing conditions ($m(R, V_b)$) and b) steady initial RS ($m(R_0, v)$), in the resistive range and for the bias voltages the physical DUT was tested.

Fig. 7 reveals the switching rate surface reproduced by the proposed model expressions (8) and the parameter values (9), (10) along with the switching rate measurements that resulted by the application of the optimizer routine on the DUT. The particularly good matching of the proposed expression (8) versus experimental results corroborates the assumptions upon which our model was structured.

## IV. MODEL INTEGRATION IN CIRCUIT SIMULATOR

A continuous and differentiable mathematical description is a mandatory requirement for any model implementation to work properly with the iterative solution methods used by computer solvers. Furthermore, for model integration into a circuit simulator with the use of the Verilog-A (VA) language, the choice of the appropriate coding approach determines the simulator's performance in efficiently simulating the model. Here, we start by producing a continuous and differentiable approximation of the proposed memristor model according to the guidelines depicted in [18], [19]. Next, we proceed in presenting the VA coding approach selected for robust simulation of the model mathematical expression

### A. Model Approximation

By inspecting the model's mathematical formulation (8), we notice that its behaviour is regulated by the discontinuous step function $stp(\cdot)$. Thus, it is replaced by its continuous sigmoid approximation $\theta_i(x) = 1/(1 + exp(-x/bi))$ (11), by properly adjusting the parameter $b_i$ for each case to facilitate simulator convergence and to keep the model's dynamics practically unaffected. The role of $b_i$ is to control the slope of the sigmoid function around the discontinuous corner points imposed by the piecewise nature of the step function. Thus,

$$stp(r_p(v) - R) \approx \theta(r_p(v) - R), \quad (12)$$

$$stp(R - r_n(v)) \approx \theta(R - r_n(v)), \quad (13)$$

$$stp(\pm v) \approx \theta(\pm v), \quad (14)$$

where $b_R = 1$ (for (12), (13)) and $b_v = 10^{-3}$ (for (14)).

The specific parameter values for $b_v$ and $b_w$ where chosen so as to retain the models dynamics practically unaffected and at the same time facilitate the simulator to produce solutions adequately fast. Further accuracy may be achieved by lowering these values at the cost though of simulation speed.

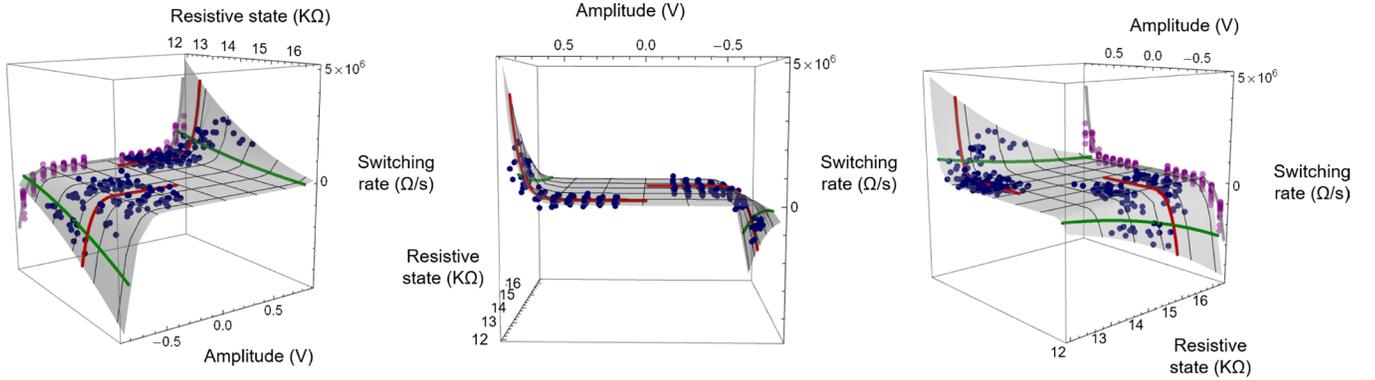

Fig. 7. Measured (blue symbols) and exhibited by the proposed model (8) 'switching surface' (gray surface) for three angles of observation. The model shows excellent agreement with measured data thus validating the stationary switching surface assumption and the considered form of the models DAE set (1), (2) in capturing device RS evolution. Red lines reproduce $dR/dt$ vs $v$ responses calculated by (8) for the RSs data was conceived ($RS0_p = 12.60K\Omega$ and $RS0_n = 14.90K\Omega$). Purple symbols with grey lines are the surface and blue point projections on the $dR-V$ plane for the refined data and the fittings also shown in Fig. 4(d). Green lines are calculated from (8) by sweeping $R$ for $v = V_b = 0.8V$.

*B. Verilog-A coding details*

In probably every SPICE memristor model published so far [10], [11], [12], [13], [20], [21] the memory effect of the memristor is modelled via a feedback controlled integrator circuit where the value of the device's state variable is represented by the voltage across a unitary capacitance which serves as an integrator of the internal state variable function. We follow a similar approach in the presented VA implementation in the sense that our model also captures RS evolution by assigning this on an internal voltage node.

What we propose is the straightforward integration of the state variable function (8) with the use of the built in VA time-domain integration operator `"idt()"`. Numerical integration may produce errors in sensitive models with extremely long time constants but this operation is far more preferable than time-domain differentiation [8]. The integral present in (8) corresponds to a recursive relationship. VA supports the calculation of a differential system that contains recursive relationships as long as the variables that express recurrence are assigned on voltage nodes or branch current flows [22]. Similar to SPICE implementations, the state variable (device RS) in our proposed implementation is assigned on an internal voltage node. By following this approach, the DAE set is integrated with the use of simulator dedicated algorithms with a plethora of available settings for the user designed to produce reliable solutions when properly adjusted. Integration in our model is executed rather fast in the Spectre simulator since all of the model expressions resulting from the smoothing procedure are continuous.

Another important point is the potential of numerical overflows imposed by the exponential functions in the sigmoid approximation kernels (12) - (14). This is dealt by using the limiting exponential function `"exp(·)"`. Function `"limexp(·)"` limits potentially overflowed values by linearizing the exponential response after an internally defined threshold [22].

The VA code for the proposed ReRAM model after applying all aforementioned modifications and specific implementation strategies (such as the integration scheme), can be found in [23]. The parameter values are defined according to the tested device (9), (10). The code is ready to use and can be compiled in any electrical circuit simulator that supports VA modules.

## V. Conclusions

In this work, we have presented a novel ReRAM model implementation method based on data from experiments and measurement analysis procedures specifically designed for capturing the switching behaviour of a typical $TiO_2$ ReRAM prototype as a function of bias voltage and device resistive state. The proposed model captures the aforementioned dependencies in a compact functional form that is able to reproduce the evolution of the device resistive state on voltage stimulation in its resistive range of operation. Our model expressions are validated by comparing measured and modelled 'switching surface' plot for the resistive region the DUT exhibits during the experiments. Furthermore, the mathematical formulation of the model is approximated as a continuous and differentiable algebraic equation set suitable for integration in circuit simulators. Finally, practical guidelines for coding the model expressions in a robust Verilog-A module are discussed.

## References

[1] D. B. Strukov, G. S. Snider, D. R. Stewart, and R. S. Williams, "The missing memristor found.," *Nature*, vol. 453, no. 7191, pp. 80–3, 2008.

[2] L. O. Chua, "Memristor—The Missing Circuit Element," *IEEE Trans. Circuit Theory*, vol. 18, no. 5, pp. 507–519, 1971.

[3] F. Bedeschi, R. Fackenthal, C. Resta, E. M. Donze, M. Jagasivamani, E. C. Buda, F. Pellizzer, D. W. Chow, A. Cabrini, G. M. A. Calvi, R. Faravelli, A. Fantini, G. Torelli, D. Mills, R. Gastaldi, and G. Casagrande, "A bipolar-selected phase change memory featuring multi-level cell storage," in *IEEE Journal of Solid-State Circuits*, 2009, vol. 44, no. 1, pp. 217–227.

[4] A. F. Vincent, J. Larroque, N. Locatelli, N. Ben Romdhane, O. Bichler, C. Gamrat, W. S. Zhao, J. O. Klein, S. Galdin-Retailleau, and D. Querlioz, "Spin-transfer torque magnetic memory as a stochastic memristive synapse for neuromorphic systems," *IEEE Trans. Biomed. Circuits Syst.*, vol. 9, no. 2, pp. 166–174, 2015.

[5] T. Prodromakis, K. Michelakis, and C. Toumazou, "Switching mechanisms in microscale memristors," *Electron. Lett.*, vol. 46, no. 1, p. 63, 2010.

[6] S. Gaba, P. Sheridan, J. Zhou, S. Choi, and W. Lu, "Stochastic memristive devices for computing and neuromorphic applications.," *Nanoscale*, vol. 5, no. 13, pp. 5872–8, 2013.

[7] A. Serb, A. Khiat, and T. Prodromakis, "An RRAM Biasing Parameter Optimizer," *IEEE Trans. Electron Devices*, vol. 62, no. 11, pp. 3685–3691, 2015.

[8] A. Serb, J. Bill, A. Khiat, R. Berdan, R. Legenstein, and T. Prodromakis, "Unsupervised learning in probabilistic neural networks with multi-state metal-oxide memristive synapses," *Nat. Commun.*, vol. 7, p. 12611, Sep. 2016.

[9] I. Gupta, A. Serb, R. Berdan, A. Khiat, A. Regoutz, and T. Prodromakis, "A Cell Classifier for RRAM Process Development," *IEEE Trans. Circuits Syst. II Express Briefs*, vol. 62, no. 7, pp. 676–680, 2015.

[10] Z. Biolek, D. Biolek, and V. Biolková, "SPICE model of memristor with nonlinear dopant drift," *Radioengineering*, vol. 18, no. 2, pp. 210–214, 2009.

[11] Y. V. Pershin and M. Di Ventra, "SPICE model of memristive devices with threshold," *Radioengineering*, vol. 22, no. 2, pp. 485–489, 2013.

[12] R. Berdan, C. Lim, A. Khiat, C. Papavassiliou, and T. Prodromakis, "A memristor SPICE model accounting for volatile characteristics of practical ReRAM," *IEEE Electron Device Lett.*, vol. 35, no. 1, pp. 135–137, 2014.

[13] Q. Li, A. Serb, T. Prodromakis, and H. Xu, "A memristor SPICE model accounting for synaptic activity dependence," *PLoS One*, vol. 10, no. 3, 2015.

[14] D. Biolek, M. Di Ventra, and Y. V. Pershin, "Reliable SPICE simulations of memristors, memcapacitors and meminductors," *Radioengineering*, vol. 22, no. 4, pp. 945–968, 2013.

[15] A. Serb, R. Berdan, A. Khiat, C. Papavassiliou, and T. Prodromakis, "Live demonstration: A versatile, low-cost platform for testing large ReRAM cross-bar arrays," in *Proceedings - IEEE International Symposium on Circuits and Systems*, 2014, p. 441.

[16] R. Berdan, A. Serb, A. Khiat, A. Regoutz, C. Papavassiliou, and T. Prodromakis, "A ??-Controller-Based System for Interfacing Selectorless RRAM Crossbar Arrays," *IEEE Trans. Electron Devices*, vol. 62, no. 7, pp. 2190–2196, 2015.

[17] R. Berdan, T. Prodromakis, A. Khiat, I. Salaoru, C. Toumazou, F. Perez-Diaz, and E. Vasilaki, "Temporal processing with volatile memristors," in *Proceedings - IEEE International Symposium on Circuits and Systems*, 2013, pp. 425–428.

[18] G. J. Coram, "How to (and how not to) write a compact model in Verilog-A," *Proc. 2004 IEEE Int. Behav. Model. Simul. Conf. 2004. BMAS 2004.*, no. Bmas, pp. 97–106, 2004.

[19] C. C. McAndrew, G. J. Coram, K. K. Gullapalli, J. R. Jones, L. W. Nagel, A. S. Roy, J. Roychowdhury, A. J. Scholten, G. D. J. Smit, X. Wang, and S. Yoshitomi, "Best Practices for Compact Modeling in Verilog-A," *IEEE J. Electron Devices Soc.*, vol. 3, no. 5, pp. 383–396, 2015.

[20] H. Abdalla and M. D. Pickett, "SPICE modeling of memristors," in *Proceedings - IEEE International Symposium on Circuits and Systems*, 2011, pp. 1832–1835.

[21] S. Kvatinsky, E. G. Friedman, A. Kolodny, and U. C. Weiser, "TEAM: Threshold adaptive memristor model," *IEEE Trans. Circuits Syst. I Regul. Pap.*, vol. 60, no. 1, pp. 211–221, 2013.

[22] Accellera, "Verilog-AMS Language Reference Manual," *http://accellera.org/images/downloads/standards/v-ams/VAMS-LRM-2-3-1.pdf*, 2009. .

[23] Messaris, I. A compact Verilog-A ReRAM model (2016). [Online]. Available: http://doi.org/10.5258/SOTON/403321